\documentclass[aps,prb,amsfonts,amssymb,epsfig,twocolumn,groupedaddress,showpacs]{revtex4} 
\usepackage{latexsym}
\usepackage{epsfig} 
\usepackage[utf8]{inputenc}
\usepackage[english]{babel}
\usepackage{amsmath}
\usepackage{color}

\usepackage{esint}
\usepackage{amsthm}

\begin{document}

\title{Edge channel mixing induced by potential steps in an integer quantum Hall system}

\author{D. Venturelli}
\affiliation{Institut NEEL, CNRS and Universit\'{e}  Joseph Fourier, Boite Postale 166, 38042 Grenoble, France\\
NEST, Scuola Normale Superiore and Istituto Nanoscienze-CNR, Piazza dei Cavalieri 7, I-56126 Pisa, Italy\\
International School for Advanced Studies (SISSA), Via Bonomea 265, I-34136 Trieste, Italy
}
\author{V. Giovannetti, F. Taddei and R. Fazio}
\affiliation{NEST, Scuola Normale Superiore and Istituto Nanoscienze-CNR, Piazza dei Cavalieri 7, I-56126 Pisa, Italy}
\author{D. Feinberg}
\affiliation{Institut NEEL, CNRS and Universit\'{e}  Joseph Fourier, Boite Postale 166, 38042 Grenoble, France}
\author{Gonzalo Usaj and C. A. Balseiro}
\affiliation{Centro At\'{o}mico Bariloche and Instituto Balseiro, Comisi\'{o}n Nacional de Energ\'{ı}a At\'{o}mica, 8400 S. C. de Bariloche and CONICET, Argentina}

\date{\today}
\begin{abstract}
We investigate the coherent mixing of co-propagating edge channels in a quantum Hall bar produced by step potentials. In the case of two edge channels it is found that, although a single step induces only a few percent mixing, a series of steps could yield 50\% mixing. In addition, a strong mixing is found when the potential height of a single step allows a different number of edge channels on the two sides of the step.
Charge density probability has been also calculated even for the case where the step is smoothened.
\end{abstract}

\date{\today}

\maketitle

\section{Introduction}
\label{intro}
When a two dimensional electron gas (2DEG) is subject to a large magnetic field, the integer quantum Hall (IQH) regime is accessed.
Here charge transport is allowed by the formation of edge-state channels, each accounting for a single quantum of conductance.
As pointed out for the first time in Refs.~\onlinecite{key-3}, this system became the prototype of a single-channel conductor with spectacular properties such as chirality and adiabatic transport, whose study fueled an enormous amount of work in the field of nanoscience\cite{key-111}. 
Recently, phase-coherence was studied and found to be preserved over rather long distances, of the order of more than 10 micrometers \cite{key-7}.
For this reason 2DEGs in the IQH regime appear to be specially suited for electronic interferometry, a very stimulating phenomenon both for basic science and for its various possible applications.
A recent breakthrough in this field has been the experimental realization of electronic Mach-Zehnder~\cite{ji,neder,litvin,roulleau,neder2} and Hanbury-Brown and Twiss~\cite{neder3} interferometers.
In these experiments electrons in the edge states loop around an annular structure mimicking the optical paths of their photonic counterparts.

Recently, a new theoretical scheme was proposed\cite{key-6} which would allow for a concatenation of several Mach-Zehnder interferometers (MZIs) in series. This new opportunity of scalability, which is not topologically possible in many of the setups experimentally developed so far, exploits the interference between adjacent edge channels with the same chirality, coupled by means of some localized potential.
Coherent mixing among co-propagating IQH channels has been investigated in recent times mainly between spin-resolved channel (induced by the spin-orbit interaction)~\cite{spinorbit}, or in connection with inelastic scattering at high chemical potential imbalance~\cite{deviatovBIAS}.
Furthermore, an intereferometer that exploits non-engineered scattering mechanisms between adiacent spin-resolved channels has been realized and successfully tested in Ref.~\onlinecite{deviatovINT}.
While the possibility of locally breaking the adiabatic transport in IQH systems has been recognized long time ago\cite{key-10}, there is now a call for a more focused study on how much adjacent cyclotron-resolved (i.e. corresponding to different Landau levels) co-propagating edge channels might be influenced by an engineered non-adiabatic potential.

In this paper we investigate the possibility of inducing coherent mixing between two co-propagating edge channels in a Hall bar due to an abrupt (non-adiabatic) potential steps along the direction of propagation.
More precisely, we calculate the inter-channel transmission probability between two co-propagating edge states induced by the potential step, which is directly connected to the conductance of the system through the Landauer-B\"uttiker current formula  -- see Eq.~(\ref{eq:100}) in the following.
The implementation of such local, short-scale potential variations is, in principle, within the experimental reach of
cutting-edge technology, for example, through: i) precise impurity implantation by means of focused ion beam~\cite{diaco}, AFM induced oxidation~\cite{shash}, cleaved-edge overgrown technique~\cite{huber}, and tunable scanning gate microscopy~\cite{heun}.

For the sake of clarity here we focus on idealized configurations. 
We first consider the case of a single potential step where two edge channels are open on its left and right hand side, finding that the channel mixing probability is pretty small even for heights of the potential step of the order of the Landau level (LL) separation $\hbar\omega_{c}$.
Moreover, in the presence of a single edge channel on both sides of the step, we find that no reflection is allowed as long as the width of the bar is larger than a few magnetic lengths.
By placing in series a number of such potential steps, though, channel mixing of the order of 50\% could realistically be achieved.
The situation changes when a single edge channel is open on the left hand side, while two channels are open on the right hand side of a potential step.
Here channel mixing can be as high as 30\% for a (single)  sharp step. 
Finally we calculate the stationary charge density in the Hall bar even in the case where the potential step in smoothed, finding indications, in all situations examined, that channel mixing persists (within the same order of magnitude) as long as the potential changes over a distance not exceeding few magnetic lengths.

All the results presented here are obtained neglecting electron-electron interaction, which is expected to be important only when a finite chemical potential imbalance is imposed between two IQH edge states and dominate the energy exchange between them in the presence of non-equilibrium electron distributions (see Refs.~\onlinecite{interacting} for different theoretical models). In the zero-bias regime, however, electron-electron interactions are not proven to play an important role, the interference pattern in MZI experiments being consistent with the single-particle theory~\cite{key-22}.

The paper is organized as follows.
In Sec.~\ref{model} we specify the system under study and we describe the numerical technique used for our calculations.
In Sec.~\ref{sec:Results} we discuss the results obtained when the abrupt step potential connects two regions characterized by the same edge filling factor (\ref{sec:equal}), and in the case of a series of such steps (\ref{res0}).
In Sec.~\ref{res2} we consider the case with one open channel on the left and two open channels on the right. 
Finally, Sec.~\ref{res3} finally focuses on the charge density probability produced by the presence of the step potential, even in the case when it is smooth.

\section{Model and numerical technique}
\label{model}

\begin{figure}
\includegraphics[scale=0.5]{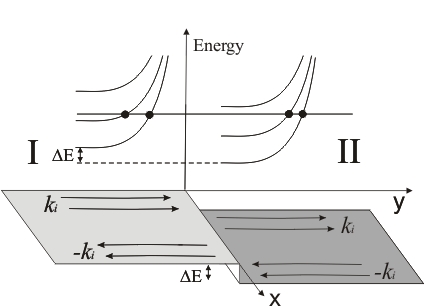}
\caption{Schematics of the set-up. A hard wall potential confines the 2DEG in the transverse direction defined by the coordinates  $x\in [-L/2,L/2]$. Along the longitudinal direction $y$ a step potential $U(y)$ is introduced to induce coherent mixing among the propagating modes. Its effect is accounted as a global energy shift between the solutions of the Schr\"odinger equation in the two regions, as pictured on the dispersion band curves of the edges drawn on the background of the figure (the horizontal line that intersects the bands indicates the Fermi energy).}
\label{Flo1}
\end{figure}

The system under investigation consists of a quantum Hall bar subjected to a sharp step-like potential $U(y)$  along the longitudinal $y$ direction (see Fig.~\ref{Flo1}), whose role is to induce scattering among otherwise independent edge-state channels. In the following we will neglect the spin degree of freedom of the electrons and consider spin-degenerate edge channels (see for example Refs.~\onlinecite{heun,heunparad2}).
The latter are determined through the  
solutions of  the  time-independent Schr\"odinger equation $H \Psi(x,y) = E\Psi(x,y)$ with the single-electron Hamiltonian (in Landau gauge) given by
\begin{equation}
\label{eq:0}
H=\frac{\hbar^{2}}{2m}[-\frac{\partial^{2}}{\partial x^{2}}+(-i\frac{\partial}{\partial y}+\frac{|e|B}{c\hbar}x)^{2}] +U(y) \;,
\end{equation}
where \emph{e }and $m$ are, respectively, the electron charge
and the effective electron mass, $B$ is the perpendicular magnetic
field.
A hard wall confinement potential that defines the
edges of the sample is assumed.
In Eq.~(\ref{eq:0}) $U(y)$ is the step potential function which
is taken to be zero for $y<0$ (region I) and constant for  positive $y$ (region II),
i.e.  $U(y)=-\Delta E\; \Theta(y)$, where $\Theta(y)$ is the Heaviside function.
Under these conditions 
the Hall bar  effectively splits  into two regions
and the resulting scattering problem  can be solved through a
\emph{mode matching} method\cite{key-5-2,key-5-22} as detailed in the following.

First we notice that in both regions the 
eigenfunctions of the Hamiltonian can be expressed as scattering
states in the y-direction (i.e. $\Psi^i\left(x,y\right)=\psi^i(x)e^{ik^iy}$, with $i=I,II$)
so that the time-independent Schr\"odinger equation reduces to 
\begin{equation}
[-\frac{\partial^{2}}{\partial x^{2}}+(k^i+\beta x)^{2}-\epsilon^{i}]\psi^i\left(x\right)=0\;,\label{eq:1}
\end{equation}
where  $i=$I, II specifies the region, $\beta=l_{B}^{-2}=\left| e\right|B/c\hbar$ is the inverse magnetic length squared, 
$\epsilon^{i}=2mE^{i}/\hbar^{2}=2\beta\left(E^{i}/\hbar\omega_{c}\right)$ is the rescaled effective energy, with $E^{\text{I}}=E$ and $E^{\text{II}}=E+\Delta E$. $E$ is defined so that the first LL corresponds to $E=\hbar\omega_c/2$, where 
$\omega_{c}=\left|e\right|B/cm$ is the cyclotron frequency.
The solutions for the transverse eigenfunction $\psi^i$ are completely specified by the magnetic field and by 
imposing hard wall boundary conditions: $\psi^i(x=-\frac{L}{2})=\psi^i(x=\frac{L}{2})=0$.
The resulting expression  is a  transcendental equation that can be expressed in analytic form in terms of parabolic cylinder functions\cite{key-5-1}.
We opt nevertheless for a numerical solution following the technical strategy detailed in the appendix of Refs.~\onlinecite{key-5-22}, {\em i. e.} discretizing Eq.~(\ref{eq:1}) in the $x$ variable.

In both regions, for a given  $E$ one can find a set of complex values  for the wave-vector $k^i$ satisfying Eq.~(\ref{eq:1}).
Those with zero imaginary part are associated to propagating longitudinal wave-functions  which correspond to the $2P_i$ edge-state channels.
$P_i$ represents the LL filling factor of region $i$, defined by the integer part of the quantity $E^i/\hbar\omega_{c}+\frac{1}{2}$ (notice that since $E^i$ differs in the two regions, $P_{\text{I}}$ and $P_{\text{II}}$ need not to coincide).  
More precisely, we can identify $P_i$ real positive solutions $\{ k_n^{i}; n=1, \cdots, P_i\}$ that describe propagating right-going channels $\{ \psi_{n}^{Ri}(x); n=1, \cdots, P_i\}$, and
$P_i$ real negative solutions $\{ -k_n^{i}; n=1, \cdots, P_i\}$  that describe propagating left-going channels $\{ \psi_{n}^{Li}(x); n=1, \cdots, P_i\}$.
Such modes are responsible for the electronic transport in the sample.
We normalize them in such a way that their current
flux is unity. This means that we impose: 
\begin{equation}
\label{eq:323}
\int_{-L/2}^{L/2} dx[\psi_{n}^{Ri}(x)\left(k_{n}^{i}+eA_{x}\right)\psi_{n}^{Ri}(x)^{*}]=1 
\end{equation}
where $A_{x}=\beta x/e$ is the only non-zero component of the vector potential in the Landau gauge. The normalization of $\psi_{n}^{Li}(x)$ follows by the symmetry of the problem that imposes
$\psi_{n}^{Li}(x)= \psi_{n}^{Ri}(-x)$ for all  $n$ and $i$.
The complex  and purely imaginary solutions, instead, are associated with evanescent eigenfunctions  $\bar{\psi}_{n}^{i}$ of the system. 
They do not contribute directly to the net electronic transport 
but are needed to guarantee the  continuity of the wave-function and of the probability current 
 when imposing the matching conditions at the boundary to the solutions\cite{key-2}, i.e. 
 \begin{eqnarray}
 \Psi^{\mathrm{I}}\left(x,y=0\right)&=&\Psi^{\mathrm{II}}\left(x,y=0\right)\;, \nonumber \\
 \partial_{y}\Psi^{\mathrm{I}}\left(x,y=0\right)&=&\partial_{y}\Psi^{\mathrm{II}}\left(x,y=0\right)\;. \label{pM}
 \end{eqnarray}
A generic solution of the Sch\"odinger equation can thus be written as follows 
\begin{eqnarray}
\Psi^{i}\left(x,y\right)&=&\sum_{n=1}^{P_i}a_{n}^{i}\psi_{n}^{Ri}\left(x\right)e^{ik_{n}^{i}y}+\sum_{n=1}^{P_i}b_{n}^{i}\psi_{n}^{Li}\left(x\right)e^{-ik_{n}^{i}y} \nonumber \\
&+&\sum_{n=1}^{Q_i}  c_{n}^{i}\bar{\psi}_{n}^{i}(x)e^{i\bar{k}_n^{i}y} \;, 
\label{eq:2}
\end{eqnarray}
where the last summation is performed over the set of the evanescent modes $\bar{\psi}_{n}^{i}$  which solve 
the Schr\"odinger equation~(\ref{eq:1})  with complex wave-vectors  $\bar{k}_{n}^{i}$. 
We stress that in principle  this last contribution should include infinitely many terms since infinite are the evanescent solutions of Eq.~(\ref{eq:1}) associated
with a given selected energy eigenvalue $E^i$.
 However,  to make the problem treatable numerically we limit the number $Q_i$ to only include those evanescent modes  $\bar{\psi}_{n}^{i}$ whose $\bar{k}_n$ lies within a finite radius from the origin of the complex plane\cite{key-5-2} (the exact number  being determined under the condition that the final result does not vary significantly if extra evanescent modes are added in the expansion -- for our simulations this corresponds to have $Q_i\simeq 20$).

Consider first the case of small $\Delta E$, {\it i. e.} where the potential step maintain the same filling factor in the two regions (i.e. $P_{\text{I}} = P_{\text{II}} =P$), and focus on the 
 scattering process associated with right-going electrons coming from the left lead  
with given mode number $j\in\{ 1, 2, \cdots, P\}$. Due to the normalization constraint of Eq.~(\ref{eq:323}), the scattering amplitudes $t_{nj}$ ($r_{nj}$) that couple such incoming mode with the transmitted (reflected) modes in the channel $n$, 
can then be directly identified with the coefficients $a_{n}^{II}$  ($b_{n}^{I}$) obtained from  Eq.~(\ref{eq:2}) while imposing the matching conditions of  Eq.~(\ref{pM}). 
The number of unknowns is given by $2(P+Q)$, since, although not entering in the scattering matrix, the coefficients relative to evanescent waves ($c_n^I$ and $c_n^{II}$) must be found.
The $2(P+Q)$ equations needed to determine them can be set by expanding the functions $\psi_n^{Ri}(x)$, $\psi_n^{Li}(x)$ and $\bar{\psi}_n^i(x)$ in the first $N/2=(P+Q)$ Fourier modes $\varphi_{n}=\sqrt{\frac{1}{L}}\sin\left(\frac{2n\pi x}{L}\right)$ as follows:
\begin{eqnarray}
\psi_{n}^{Ri}(x)=\sum_{j=1}^{N/2}\alpha_{nj}^{i}\varphi_{j}(x)~~~\mbox{for}~1\le n\le P\;, \\
\psi_{n}^{Li}(x)=\sum_{j=1}^{N/2}\beta_{nj}^{i}\varphi_{j}(x)~~~\mbox{for}~1\le n\le P \;,\\
\bar{\psi}_{n}^{i}(x)=\sum_{j=1}^{N/2}\gamma_{nj}^{i}\varphi_{j}(x)~~~\mbox{for}~1\le n\le Q \;,
\label{eq:5}
\end{eqnarray}
the coefficients $\alpha_{nj}^{i}$ corresponding to right-going modes, $\beta_{nj}^{i}$ to left-going modes, and $\gamma_{nj}^{i}$ to evanescent modes.
At the end of the simulation we check that the number of Fourier Modes used in the expansion is sufficient to properly describe all propagating, oscillatory damped and evanescent modes that contribute appreciably to the scattering matrix.
By multiplying by $\varphi_{l}$ and integrating over $x$, the above expressions can be recasted in the following $N\times N$ matrix equation:
\begin{eqnarray}
\left(\begin{array}{c}
\sum_{n}^{P}\left(a_{n}^{I}\vec{\mathbf{\alpha}}_{nl}^{I}-a_{n}^{II}\vec{\mathbf{\alpha}}_{nl}^{II}\right)\\\\
\sum_{n}^{P}\left(k_{n}^{I}a_{n}^{I}\vec{\mathbf{\alpha}^{I}}_{nl}-k_{n}^{II}a_{n}^{II}\vec{\mathbf{\alpha}}_{nl}^{II}\right)\end{array}\right)= \nonumber \\
=\left(\begin{array}{ccccc}
\mathcal{B}^{II} & -\mathcal{B}^{I} & &\mathcal{G}^{II} & -\mathcal{G}^{I} 
\\  \\
\tilde{\mathcal{B}}^{II} & -\tilde{\mathcal{B}}^{I} && \tilde{\mathcal{G}}^{II} & -\tilde{\mathcal{G}}^{I}\end{array}\right)
\left(\begin{array}{c}
\vec{b}_{n}^{II}\\
\vec{b}_{n}^{I}\\
\vec{c}_{m}^{I}\\
\vec{c}_{m}^{II}\end{array}\right)
\label{eq:6}\end{eqnarray}
where for $i= I, II$, $\vec{\alpha}_{nl}^{i}\equiv (
\alpha_{n1}^{i} , \alpha_{n2}^{i},  \dots, \alpha_{nN}^{i})^{T}$, $\vec{b}_{n}^{i}\equiv(
b_{1}^{i} ,b_{2}^{i} , \dots , b_{P}^{i})$, $\vec{c}_{n}^{i}\equiv (
c_{1}^{i} , c_{2}^{i} , \dots , c_{Q}^{i})$, and $\mathcal{B}^{i}$, $\mathcal{G}^{i}$ denote the matrices containing
the Fourier coefficients, namely $(\mathcal{B}^{i})_{nl}\equiv \beta_{nl}^{i}$ and $\mathcal({G}^{i})_{nl}\equiv \gamma_{nl}^{i}$ respectively, while $\tilde{\mathcal{B}}^i$ and $\tilde{\mathcal{G}}^{i}$ denote the
matrices of elements $(\tilde{\mathcal{B}}^{i})_{nl}\equiv k_{n}^{i}\beta_{nl}^{i}$ and $(\tilde{\mathcal{G}}^{i})_{nl}\equiv k_{n}^{i}\gamma_{nl}^{i}$.
This linear problem can be solved numerically so that the resulting coefficients
allow a full reconstruction of the wave-function in all regions through Eq.~(\ref{eq:2}).
The same analysis holds when $P_{\text{I}} \neq P_{\text{II}}$ with the only important requirement that the linear system in Eq.~(\ref{eq:6}) is determined, i.e. that $P_{\text{I}} + Q_{\text{I}} \equiv P_{\text{II}} + Q_{\text{II}}$. An example of such configuration is presented in Sec.~\ref{res2}  where we assumed $P_{\text{I}}=1$ and 
$P_{\text{II}}=2$. 

To conclude the section we mention that the conductance $G$ of the system is determined, according to the Landauer-B\"uttiker scattering theory~\cite{key-3}, by the expression
\begin{equation}
G=\frac{2e^2}{h} \sum_{n=1}^{P_{II}} \sum_{j=1}^{P_I} |t_{nj}|^2 ,
\label{eq:100}
\end{equation}
valid in limit of small voltages and zero temperature.

\section{Results}
\label{sec:Results}
In this section we shall discuss the results obtained for the scattering amplitudes in the case of a Hall bar with either one or two open edge channels.

\subsection{Two regions with equal filling factor}
\label{sec:equal}

\begin{figure}[h]
\includegraphics[scale=0.75]{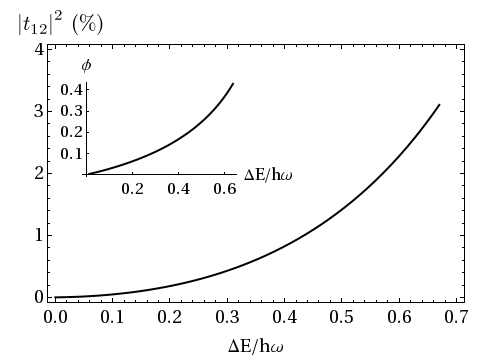}
\caption{Channel mixing probability $|t_{12}|^2$ percentage, for the case $P_I=P_{II}=2$, as a function of the height of the potential step $\Delta E$.
In the inset: scattering phase shift as a function of the potential step height for a single edge channel.}
\label{CnMix1}
\end{figure}

Let us now consider the case of two edge channels ($P_I=P_{II}=2$) on each side of the step potential, aiming at evaluating the channel mixing probabilities $\left|t_{12}\right|^{2}$ and $\left|t_{21}\right|^{2}$ representing the probability for transmission from inner (2) to outer (1) edge and vice-versa, respectively (see Fig.~\ref{Flo1}).
By setting $L=6.7l_B$, where $l_{B}=\beta^{-\frac{1}{2}}$ is the magnetic length, we make sure that the reflection probabilities are negligible. More precisely, fixing the energy of the incoming electrons at $1.7\hbar\omega_{c}$ above the first LL, we found that the only non-vanishing, though very small, reflection coefficient is $\left|r_{22}\right|^{2}\sim 10^{-3}$.
In Fig.~\ref{CnMix1} the channel mixing probability $\left|t_{12}\right|^{2}$ is plotted as a function of the potential barrier height $\Delta E$ in units of $\hbar\omega_c$:
$|t_{12}|^2$ increases monotonically with increasing $\Delta E$, taking a value of the order of few percent only for a step potential as high as $0.7\hbar\omega_{c}$ (note that, due to the non-zero reflection probability, $|t_{21}|^2$ slightly differs from $|t_{12}|^2$).

It is worth mentioning that,   in the limit of small step height $\Delta E \ll\hbar\omega_{c}$,
 an  analytical estimation of  $t_{12}$ is possible. For instance 
assuming a potential of the form $U\left(y\right)=-\Delta E\Theta\left(y\right)e^{-y/\mathcal{L}}$ while taking the limit $\mathcal{L}\longrightarrow\infty$, 
one can verify that, up to a phase factor, the channel mixing amplitude $t_{12}$ can be approximated to the first order in $\Delta E$ (Born approximation~\cite{VENTRA}) as:
\begin{equation}
t_{12}
=\frac{1}{\sqrt{\mathcal{N}_{12}}}\frac{\Delta E}{k_{1}^{I}-k_{2}^{II}}
\int dx\; \psi_{k_{1}}^{I}\left(x\right)\psi_{k_{2}}^{* II}(x) \;,
\label{eq:pertTheory}
\end{equation}
where
\begin{eqnarray}
\mathcal{N}_{12}=\left|
\int_{-L/2}^{L/2}dx |\psi_{k_{1}}^{I}(x)|^{2}(\beta x+k_{1}^{I}) \times \right. \nonumber\\ \left.
 \int_{-L/2}^{L/2}dx'|\psi_{k_{2}}^{II}(x')|^{2}(\beta x'+k_{2}^{II})
\right|\;,
\end{eqnarray}
is the normalization factor that ensures the unitarity of the scattering matrix. 
We checked that the curve reported in Fig.~\ref{CnMix1} is fitted by the formula (\ref{eq:pertTheory}) close to the origin.

As a check we also consider the case of a single edge channel ($P_I=P_{II}=1$).
Here we have verified that the reflection probability $|r_{11}|^2$ is negligible, within the numerical accuracy, as long as $L$ is greater than 6.5 $l_{B}$.
Current conservation therefore implies that one can write $t_{11}=e^{-i\phi}$.

The inset of Fig.~\ref{CnMix1} shows the phase $\phi$ in radians as a function of the potential step height $\Delta E$ in units of $\hbar\omega_c$. The energy of the impinging electrons $E$ is set to $0.8\hbar\omega_{c}$ (i.e.  $0.3\hbar\omega_{c}$ above the first LL).
The phase shift $\phi$ increases monotonically nearly reaching the value $\pi/8$ for the highest step considered.

\subsection{Series of potential steps}
\label{res0}

A possible strategy to achieve a channel mixing of the order of 50\% is to place several potential steps in series.
This is in principle possible by using nanopatterning techniques to realize a sequence of top gates.
Assuming a typical magnetic lengths of about 10 nm, a few tens potential steps could be obtained over a length of some microns.

\begin{figure}
\includegraphics[scale=0.7]{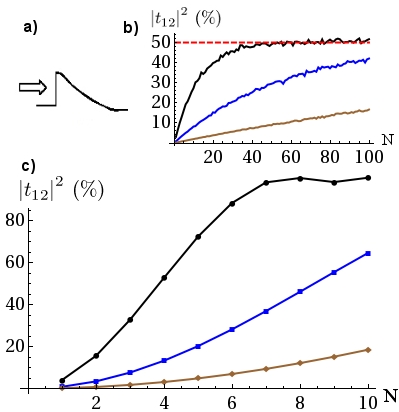}
\caption{a) Single potential step followed by an adiabatic tail. b) Averaged channel mixing probability $\left|t_{12}\right|^{2}$ as a function of $M$ for different potential heights (blue: $0.72\hbar\omega_{c}$, purple: $0.4\hbar\omega_{c}$, brown: $0.2\hbar\omega_{c}$) assuming random phases $\phi_i$ accumulated between the steps. Numerical error on unitarity of the S-matrix might induce variations of the order of 1\%. The curves represent the average over 2000 random configurations. c) Channel mixing probability $\left|t_{12}\right|^{2}$ as a function of $M$ for different potential heights (same color code as for panel b) assuming that each individual phase-adjusting gate is tuned to maximize the mixing. }
\label{Flo:schem}
\end{figure}

A simple evaluation of the channel-mixing transmission probability can be done by assuming that, after the sharp step, the potential smoothly goes to zero (see Fig.~\ref{Flo:schem}a).
In doing so, after the mixing occurring at a potential step, the electrons in the two channels freely propagate along the potential tail to the next potential step accumulating a relative phase.
Once suppressed all reflections due to the large separations between steps, the total transmission matrix $t(M)$ of a series of $M$ steps is (up to a global phase) the product of the transmission matrices of the individual steps (of height $\Delta E_i$) plus tails, which include the phase $\phi_i$ accumulated while propagating past the step $i$:
\[
t(M)=\prod_{i=1}^{M}\left(\begin{array}{cc}
t_{11}\left(\Delta E_{i}\right)e^{i\phi_{i}} & t_{12}\left(\Delta E_{i}\right)e^{-i\phi_{i}}\\
t_{21}\left(\Delta E_{i}\right)e^{i\phi_{i}} & t_{22}\left(\Delta E_{i}\right)e^{-i\phi_{i}}\end{array}\right) \,.
\]
The phase $\phi_{i}$  depends both on the details of the adiabatic tail of the step and on the distance $x_i$ between the steps.
It turns out that even a few steps can increase dramatically the channel mixing probability $|t_{12}(M)|^2$ and that the latter, due to interference effects, very much depends on the set of phases $\{\phi_{i}\}_{i=1,M}$.
For example, 50\% mixing can be achieved with four potential steps of height $\Delta E\simeq0.72\hbar\omega_{c}$, or with 10 potential steps of height $\Delta E\simeq0.4\hbar\omega_{c}$.
The control of the phases $\phi_{i}$, in order to tune the channel mixing, can be obtained by placing lateral finger gates in the region of the tail of the potentials.
The role of these additional gates is  to modify the lateral confinement potential in such a way to alter the distance $x_i$ traveled by the electrons propagating between two steps.
Indeed, due to the large difference $(k^i_{1}-k^i_{2})$, even a small variation of $x_i$ (of the order of 1/10 of the magnetic length) results in a very significant variation of phase difference between the modes $\phi_{i}=(k^i_{1}-k^i_{2})x_i\simeq1$.
In Fig.~\ref{Flo:schem}c the maximum (over $\phi_i$) channel mixing probability $|t_{12}(M)|^2$ (obtained numerically) is plotted as a function of the number of potential steps for three different values of step height, namely $0.2\hbar\omega_c$, $0.4\hbar\omega_c$ and $0.72\hbar\omega_c$.

It is interesting to consider the situation where the phase differences $\phi_i$ are not controlled and take random values.
In this case for every $M$ one can average the channel mixing probability over a given number of configurations of the set $\{\phi_{i}\}_{i=1,M}$, with $\phi_{i}\in [0,2\pi]$.
In Fig.~\ref{Flo:schem}b we plot $|t_{12}(M)|^2$ averaged over 2000 configurations for different values of step height (the same as for Fig.~\ref{Flo:schem}c).
We notice that equilibration (50\% mixing) is reached for a large enough $M$.

\subsection{Two regions with different filling factor}
\label{res2}

An alternative possible  strategy for obtaining a significant channel mixing consists in fixing $P_I=1$ and setting $\Delta E$ large enough so that in region II two edge channels are open ($P_{II}=2$).
In this case the incoming electrons will be split between the two edge channels available in region II, according to the values of the transmission amplitudes $t_{21}$ and $t_{11}$.

\begin{figure}
\includegraphics[scale=0.66]{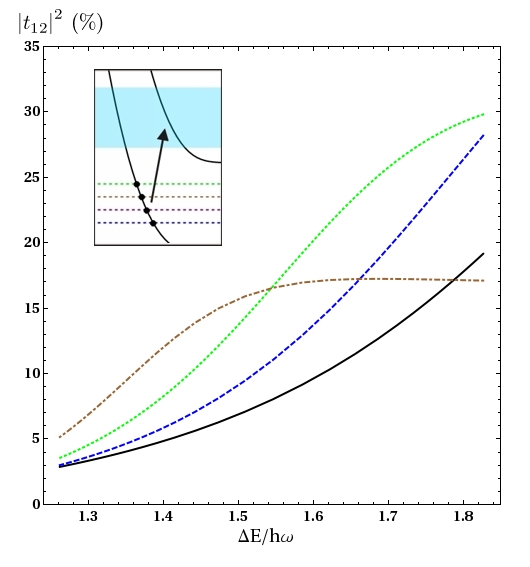}
\caption{Channel mixing probability  $\left|t_{12}\right|^{2}$ in the case where $P_I=1$ and $P_{II}=2$ for four different values of energy of the incoming electrons (pictured as dashed lines in the inset) as a function of the potential step height $\Delta E$, which spans the energies indicated on the shaded area on the inset}
\label{Flo:4-4}
\end{figure}

In order to qualitatively characterize the effect, Fig.~\ref{Flo:4-4} shows the probability $|t_{21}|^2$ for some indicative values of incident energy $E$ spread all over the energy gap, and as a function of the energy step $\Delta E$.
For all the curves channel mixing exceeds 15 \%, reaching about 30 \% for $E=1.6\hbar\omega_{c}$ and $E=1.7\hbar\omega_{c}$.
\\
We emphasize that this setup might be used to create the initial coherent superposition of wave-packet on the two edge channels which are needed for the interferometer of Ref.~\onlinecite{key-6}.

\section{Electron probability density}
\label{res3}

\begin{figure}
\includegraphics[scale=0.8]{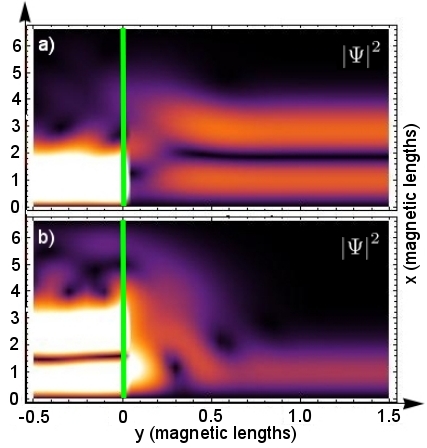}
\caption{Charge probability density color plot of the edge states in the case where $P_I=2$ and $P_{II}=2$ with a sharp step potential. Vertical lines represent the position of the step potential. Electrons are injected from region $I$ in channel 1 (a) and channel 2 (b). For the sake of clarity, only the contribution to the wave-function relative to $t_{12}$, for panel (a), and relative to $t_{21}$, for panel (b), are retained.}
\label{Flo:4-7}
\end{figure}

In this section we address the electron probability density $|\Psi (x,y)|^2$ in the case of two edge channels in region $II$ ($P_{II}=2$).
In Fig.~\ref{Flo:4-7} the density $|\Psi (x,y)|^2$ is plotted in the case of a sharp step potential with $P_I=2$ where electrons are injected from region $I$ in channel 1 (a) and channel 2 (b).
Vertical lines represent the position of the potential step ($y=0$), so that region $I$ is on the left hand side and region $II$ is on the right hand side.
Bright areas in region $I$ correspond to the high probability density of incoming electrons exhibiting, in the transverse $x$-direction, one lobe, for injection from channel 1, and two lobes, for injection from channel 2.
In region $II$ the probability density relative only to the transmitted electronic wave functions with channel mixing is plotted, i.e. the contribution to the wave functions due to the amplitudes $t_{11}$ (for panel (a)) and $t_{22}$ (for panel (b)) has been subtracted for clarity.

\begin{figure}
\includegraphics[scale=0.8]{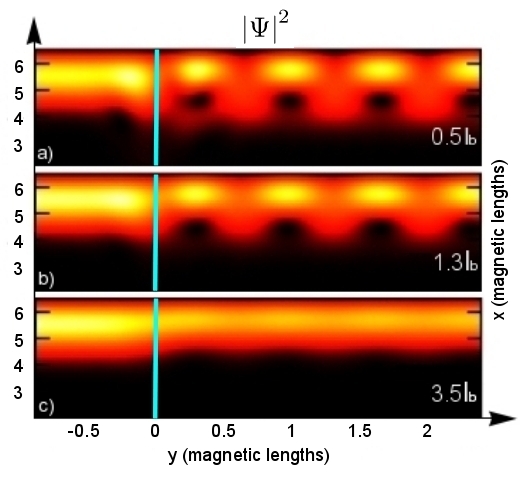}
\caption{Charge probability density color plot of edge states in the case where $P_I=1$ and $P_{II}=2$ with a smooth step potential characterized by a width $d$ indicated in figure. Vertical lines correspond to the center position of the step potential. Panel (a), (b) and (c) are relative to, respectively, $d$=0.5, 1.3 and 3.5 magnetic lengths.
For $d=0.5l_{B}$ the plot is indistinguishable from the one obtained with a sharp step.}
\label{Flo:4-2}
\end{figure}

Up to now we have considered the ideal situation in which the step potential is sharp.
Now we address the effect of the smoothening of the step and describe the cross-over to the adiabatic regime occurring when the potential varies over a length which is larger than the magnetic length.
For these calculations we make use of a tight-binding model where the wave-function is computed by means of the recursive Green's functions
technique, applied successfully in other contexts\cite{key-9}. 
Numerical simulations are performed by replacing the sharp step with a potential of the form $U(y)=-\Delta E/(e^{y/d}+1)$, where $d$ is the characteristic length (width) of the potential.
In Fig.~\ref{Flo:4-2} contour plots of the probability density are shown when electrons are injected from the left in channel 1 for three different values of $d$, namely $d=0.5l_{B}$ (a), $d=1.3l_{B}$ (b) and $d=3.5l_{B}$ (c).
Vertical lines represent the center position of the smooth step potential.
Figure~\ref{Flo:4-2}(a) shows that, for $d=0.5l_{B}$, there are beatings on the right hand side of the barrier which correspond to the coherent superposition of electronic waves over the two edge channels (the period of the oscillations corresponds to $2\pi$ divided by the difference of the wave-vectors of the two outgoing modes, as expected).
Such beatings are progressively suppressed as the barrier becomes smoother, eventually disappearing for $d=3.5l_{B}$ (see Fig.~\ref{Flo:4-2}c), when the edge channel injected from region $I$ is totally transmitted to region $II$ without mixing.
It is worthwhile noting that the plot relative to $d=0.5l_{B}$ is indistinguishable from the plot relative to a sharp edge.
All simulations that we have performed confirm the picture of a crossover from the channel mixing situation to the adiabatic regime, reached when the potential step varies over a scale of a few magnetic lengths.

\section{Conclusions}
In this paper we have investigated the edge channel mixing due to steps potentials in a 2DEG in the integer quantum Hall regime.
Coherent mixing can be linked to the zero-bias linear conductance of each individual channel by the Landauer-Buttiker formalism.
Recent experiments indicate that localized scattering might couple drastically cyclotron-resolved edge channels\cite{heunparad2}, and non-adiabatic engineered potentials are thought to be the key for the implementation of scalable electronic interferometers implemented by using IQH edge channels, according to Ref.\onlinecite{key-6}.

In the case of a single sharp step we have found that, in the presence of two edge channels on each side of the step, the channel mixing probability cannot be larger than a few percent.
Channel mixing, though, can be substantially enhanced by putting in series a small number of steps.
More precisely, 50\% mixing can be already be reached with 4 steps of large height, provided that one can control the phase accumulated by the electrons propagating between two consecutive steps.
A quite large mixing can also be attained if the height of the steps is large enough to allow a single channel only on its right hand side.
In the last section, we have finally addressed the effect of the step potential on the electron density probability even in the case where the step is smooth.
Our findings suggest the possibility of employing engineered breaking of the adiabatic transport regime of IQH edge channels as a tool to induce scattering among
otherwise independent propagating modes, which can be relevant in the characterization of the coherent transport.
As a future direction we plan to extend our results investigating the role of interactions, whose effect is not negligible in regimes where the confinement potential is smooth and large bias is applied.

This work was by the Italian MIUR under the FIRB IDEAS project RBID08B3FM, by the "Universita Italo Francese/Université
Franco Italienne" (UIF/UFI) under the Program VINCI 2008 (chapter II), by the EU-project NANOCTM and by the ECOS-Sud program of
French Government.

\end{document}